Title: Spontaneous symmetry breaking and pattern formation of organoids

Keisuke Ishihara[1,2,3,*], Elly M. Tanaka[4,*]

1: Max Planck Institute of Molecular Cell Biology and Genetics, Dresden, Germany
2: Max Planck Institute for the Physics of Complex Systems, Dresden, Germany
3: Center for Systems Biology Dresden, Dresden, Germany
4: Research Institute of Molecular Pathology, Vienna, Austria
*: Correspondence ishihara@mpi-cbg.de and elly.tanaka@imp.ac.at

Abstract

Recent 3D organ reconstitution studies show that a group of stem cells can establish a body axis and acquire different fates in a spatially organized manner. How such symmetry breaking happens in the absence of external spatial cues, and how developmental circuits are built to permit this is largely unknown. Here, we review spontaneous symmetry breaking phenomena in organoids, and hypothesize underlying patterning mechanisms that involve interacting diffusible species. Recent theoretical advances offer new directions beyond the prototypical Turing model. Experiments guided by theory will allow us to predict and control organoid self-organization.

# Introduction

Pattern formation is a fundamental aspect of developmental biology. To produce the diverse cell types of the organism, cell fate specification must occur in a spatially organized manner within a cellular population. The classic model for tissue patterning posits an asymmetric distribution of a diffusible signaling molecule, whose local concentration is interpreted by cells as positional information[1]. Numerous studies have proposed how such morphogen gradients are established and interpreted [2]. A common assumption for morphogen gradient formation is a localized source of the diffusible signal. In contrast, Alan Turing proposed that two interacting diffusible species is sufficient for "spontaneous" pattern formation starting from uniform initial conditions [3]. The theory famously predicts that a pair of "local activation with long-range inhibition" forms stripes and dot patterns [4]. The concept of diffusion-driven instability has been an influential model of how symmetry, or the uniformity, can be broken in the absence of external spatial cues.

While symmetry breaking has been studied at the level of single cells [5], we know much less about symmetry breaking of cellular populations, namely the spontaneous formation of a body axis. In living organisms, a tissue is spatially and temporally placed in the context of other tissues and prior developmental events. Thus, tissue symmetry is often broken before patterning begins. As an example, the nodal flow-induced left-right symmetry breaking, originates from the structural chirality of cytoskeletal molecules and the outcome is deterministic with respect to the pre-existing anterior-posterior axis [6]. Tracing back the origin of asymmetry during development, we arrive at early embryos. In systems such as frog eggs and *C. elegans* eggs, the site of sperm entry defines the body axes at the single cell stage. Fruit fly oocytes have differentially distributed maternal factors, so symmetry is broken prior to fertilization. Perhaps, mammalian blastocysts [7,8] and chick embryos [9,10] are the only accessible *in vivo* system where radial symmetry breaking likely occurs spontaneously within a group of cells. Thus, we have had limited means to study spontaneous symmetry breaking mechanisms that may operate at various tissues at later stages of development.

What potential do undifferentiated cells have to pattern themselves? How can we study spontaneous pattern formation and symmetry breaking in a context removed from the spatial bias of external factors? A powerful approach is to use pluripotent stem cells to reconstitute tissues *in vitro*. Early observations from embryoid bodies demonstrated that an apparently uniform aggregate of stem cells elaborates tissue scale organization including multiple germ layers [11]. In the past decade, organoid methods have drastically improved the reproducibility of stem cell self-organization phenomena to the extent that researchers can now study organ development by reconstitution (reviewed in [12-14]). Organoid methods have demonstrated that embryonic stem (ES) cells and adult stem cells have the intrinsic ability to spontaneously self-organize into a variety of tissues.

Today, we are in a position to use organoids to uncover the basic principles of stem cell self-organization [15]. While such collective behavior should ultimately be understood as the successive interplay between mechanical and chemical processes, we will benefit from studying problems where the two can be uncoupled. In this mini-review, we discuss three organoid models with minimal but sufficient complexity to investigate spontaneous symmetry breaking (Box 1). In all cases, an initial population of a single cell type is grown under uniform 3D conditions, but differentiates in a spatially organized manner. Thus, these systems represent ideal conditions to experimentally test Turing's ideas for pattern formation. Our particular focus will be to ask if existing theoretical models are sufficient to explain the current observations, and, if not, what types of novel theory and experiments are anticipated.

## Box 1. Spontaneous symmetry breaking as a hypothesis

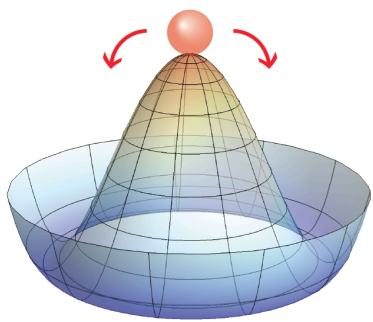

In modern physics, spontaneous symmetry breaking is formally defined when the Lagrangian ("the rules for how the system evolves") preserves the symmetry, but the lowest energy solutions are asymmetric. We can intuitively understand this concept with the famous example of Goldstone's "Mexican hat" potential. If a ball is placed at the peak of such potential, it can roll down and settle to any position within the circular well, breaking radial symmetry. Physicists have used the concept of spontaneous symmetry breaking to understand how materials become ferromagnetic and predict the mass of elementary particles. Admittedly, there are several challenges in applying this formal definition to biological systems, which have many hidden variables and are characterized, at best, as quasi-steady state conditions. Nevertheless, we believe that spontaneous symmetry breaking can serve as a guiding hypothesis for us to investigate biological self-organization.

## Rostral-caudal patterning in cortical organoids

Cortical organoids recapitulate early forebrain development starting from an aggregate of embryonic stem cells [16,17], and show spontaneous rostral-caudal patterning [18] (Fig. 1, left). The earliest sign of symmetry breaking is the polarized expression of rostral marker Six3, which is accompanied by anti-correlated expression of Fgf5. FGFR/MAPK signaling inhibits Six3+ rostral specification, while locally inducing Wnt ligand expression. The resulting Wnt activity gradient consolidates the rostral-caudal pattern marked by a Six3+ half and Irx3+ half. The authors suggest the "loss of Fgf signaling" and "increased Wnt signaling" as the molecular mechanisms underlying Nieuwkoop's "activation-transformation" model [19] for rostral-caudal neuroepithelial patterning [18]. The patterning outcome of cortical organoids can be described as "tissue polarization" where the tissue is divided in two contiguous halves of different fates. This is qualitatively different from the periodic patterns, the hallmark prediction of Turing-type mechanisms. One possible explanation is that the 300-400 μm tissue diameter coincides with half the wavelength of the underlying Turing type mechanism. Alternatively, is this a result of a different class of patterning mechanism that robustly generates a polarized tissue without ever producing periodic patterns?

This motivates us to review an emerging class of theoretical models that have been developed to explain "cell polarization", or the spontaneous symmetry breaking of single cells such as bud site selection in yeast, polarization of motile cells, and polarization of the *C. elegans* single cell. We propose that the core concepts of "cell polarization" may be applicable to organoid-scale patterning. To explain cell polarization phenomena, Edelstein-Keshet and colleagues introduced the so called "wave-pinning" model consisting of two interacting diffusible species with mass conservation [20-22]. The interconversion between the "active" and "inactive" forms are defined by a bistable reaction term. An additional assumption is that the active species diffuses slowly, but the inactive species diffuses quickly and is always spatially uniform. Starting from a uniform "inactive" state, the model predicts that sufficiently large initial fluctuations leads to a polarized pattern, where most of the active species is found on one side of the cell. It has been debated whether this represents a new class of diffusion-driven instability that is distinct from conventional Turing mechanisms, or a special limit of Turing model with subcritical bifurcation [23-26]. Perhaps, a more practical approach to understand spontaneous symmetry breaking may come from the comparative analysis based on the topology of their bifurcation diagrams [23].

What does cell polarization teach us about tissue polarization? While it is not obvious what the conserved quantity for tissue polarization may be, wave-pinning models direct us to one interesting limit in the reaction-diffusion framework. In particular, the polarization model makes a characteristic prediction: Assuming that the conserved quantity scales in larger tissues, the polarized pattern shows perfect scaling behavior. Depending on the initial distribution of fluctuations, multi-peak patterns could appear, but without any intrinsic length scale. Variants of the wave-pinning model robustly achieve single peaked patterns [27]. Experimentally, the wave-pinning model may be distinguished from Turing-type models, which predict that the number of repeating patterns with fixed length scale will increase proportionally with tissue size. For cortical organoids, theoretical models of tissue polarization motivate specific experiments that ask the effect of system size on patterning.

## Anterior-posterior patterning in 3D gastruloids

In the 3D gastruloid method, an aggregate of mouse embryonic stem cells gives rise to the three germ layers in a spatially organized manner [28,29]. At one end of the 200 μm diameter tissue, a population of cells start to express the posterior tail bud marker T/Brachury cells followed by axis elongation reminiscent of a gastrulating embryo (Fig. 1, center). Pharmacological activation of the Wnt pathway is not essential for localized Wnt pathway activation and T/Bra expression [28], but significantly enhances the robustness of the symmetry breaking outcome [30]. The Nodal pathway is absolutely required for T/Bra [30]. While *in vivo* studies underscored the importance of the trophoectoderm, an extra-embryonic tissue, for establishing the anterior-posterior axis, gastruloids seem to employ the same symmetry breaking circuit found *in vivo,* with a different activation method resulting in the surprisingly robust outcome. Strikingly, tissues larger than the optimal size result in the increased frequency of multiple T/Bra protrusions [29]. This outcome already favors a Turing-type patterning mechanism, where the number of periodic patterns scale proportionally to tissue length. In such scenario, Wnt and Nodal are candidate diffusible ligands that form periodic concentration profiles. However, one immediate issue is why ubiquitous Nodal ligands are sufficient to rescue the patterning in a *Nodal* mutant gastruloid [30].

This conflict originates from a predominant concept in pattern formation: "Local activation with long-range inhibition." For a system consisting of two interacting diffusible species, the necessary and sufficient condition for pattern formation is often called the differential diffusivity requirement [4]. While numerous combination of molecules/pathways have been proposed as the Turing pairs that satisfy this requirement (e.g. Nodal/Lefty [31], Wnt/Dkk [32], BMP/Noggin, Fgf/BMP, Shh/Fgf [33], reviewed in [34]), several theoretical studies have challenged its absolute requirement for generalized pattern formation networks [35-37], specifically, for networks that assume three or more species with at least one immobile, cell autonomous node. The cell autonomous node can be interpreted as the computation that occurs inside a cell, for example, by transcription factors. Most recently, a mathematical tool was developed to systematically screen large networks for conditions that support Turing patterns [38,39]. Using modern computer algebra, Turing's original approach of linear stability analysis was scaled for larger networks. In certain networks with two diffusible and one immobile species, pattern formation occurred with equally diffusing signals and even for any combination of diffusivities. As realistic pattern formation mechanisms will likely involve more species than a pair of activator-inhibitor molecules, demonstration of differential diffusivity may not be necessary and effort should be directed to identify the minimal number of nodes and the topology of their interactions in a patterning network.

In light of this theoretical work, future models for T/Bra pattern formation in gastruloids should likely include an immobile node in a Turing type model. The minimal network may additionally incorporate Wnt and Nodal as the diffusible cues, which topologically resembles the three component Bmp-Sox9-Wnt Turing network for digit patterning [40]. There, titration of BMP and Wnt inhibitors, and their combinations provided strong support for model validation. Similar experiments might provide an explanation to how Nodal and Wnt ligand levels affect T/Bra expression patterns and axial protrusions in 3D gastruloids.

## Dorsal-ventral patterning in neural tube organoids

Neural tube organoids are derived from single mouse ES cells grown in 3D scaffolds, and harbor a single fluid-filled lumen [41,42]. While the default positional identity corresponds to the dorsal midbrain, treatment with all-trans retinoic acid (RA) posteriorizes the tissue to hindbrain levels and simultaneously induces the establishment of a dorsal-ventral axis (Fig. 1, right). The earliest sign of symmetry breaking is the localized expression of the floor plate marker FoxA2, which is followed by Shh expression. Wnt and BMP pathways play an inhibitory role on FoxA2 expression [43] consistent with their roles as dorsal cues in the neural tube *in vivo* [44]. Shh pathway inhibition decreases the occurrence of FoxA2+/Shh+ cellular populations, but only for prolonged inhibition [41]. Thus, retinoic acid is likely activating the Shh-FoxA2 positive feedback loop through induction of FoxA2. This is in contrast to the *in vivo* sequence in amniotes where the notochord derived Shh ligand activates FoxA2 in the neural tube.

The small size of neural tube organoids, growing from 50 to 100 μm, may allow us to observe symmetry breaking dynamics at single cell resolution. In our preliminary live imaging experiments, FoxA2 expression show considerable variability among neighboring cells at the earliest time points. Later, multiple clusters of FoxA2 expressing cells are observed within the same tissue, but such a situation is often resolved by only one cluster retaining FoxA2 expression. Cell movement is present but limited, arguing against a scenario solely based on sorting of FoxA2+ cells in a differential adhesion-type mechanism. Thus, we hypothesize a competition mechanism where FoxA2+ clusters mutually inhibit FoxA2 expression via a diffusible molecule. The hypothetical inhibitor may be secreted into the lumen shared among all cells in the tissue. Within a FoxA2 cluster, there is likely a FoxA2 positive feedback mechanism, which may be cell autonomous (e.g. FoxA2 transcriptional auto-regulation[45]) and/or non-autonomous (e.g. FoxA2+ cell proliferation, cell-cell contact mediated signaling). In such scenarios, what is the role of stochastic FoxA2 transcription in symmetry breaking? How would increasing the tissue size affect the FoxA2 patterning outcome? These ideas are starting points to construct and experimentally test simple models for FoxA2 pattern formation.

## Outlook

We have reviewed three examples organoid models that enable the study of spontaneous symmetry breaking and pattern formation in the absence of spatial cues. Together with several other models [46-50], these experimental system will help us understand how stem cells self-organize. To this end, we expect theoretical models of reaction-diffusion systems to guide specific quantitative experiments. We conclude by discussing two general issues.

First, increasingly detailed observation of pattern formation phenomena prompt us to explore a wider spectrum of modeling approaches. While continuum descriptions often provide an intuitive understanding of specific scenarios, we must question the validity and sufficiency of the approximation. Thus, it will become increasingly important to compare continuum models with those that treat cells as discrete, replicating entities. In the latter approach, cell-cell communication mechanisms beyond diffusible signals such as coordination through planar cell polarity or Notch-type signaling are represented more easily. One surprising observation in cortical organoids is that inhibition of Wnt antagonists leads to an intermixed, salt and pepper-like pattern of rostral and caudal fates [18]. Continuum models fail to explain such pattern that alternates at the single cell length-scale, motivating theoretical models that describe individual cells. As another example, neural tube organoids consist of a mere ~100 cells at the onset of FoxA2 expression, whose level varies from cell to cell. Further, cells undergo several rounds of division over the course of 48 hours before symmetry is clearly broken. The stochasticity of FoxA2 induction, and clone size dynamics challenge the explanatory power of deterministic, continuum models. Thus, improved experiments will continue to inspire novel theoretical directions.

The second issue concerns the robustness of symmetry breaking phenomena. Tissue engineering applications require a higher standard of robustness, or reproducibility, of the patterning outcome than what 3D organoids currently achieve. Interestingly, 2D micropattern methods that do not break radial symmetry [51-54] show significantly more robust patterning outcomes, leading to a conjecture: Is there an intrinsic trade-off between spontaneous symmetry breaking and robustness? If such trade-off exists, approaches to spatially bias the symmetry breaking circuit by external chemical gradients will become increasingly important [55]. Otherwise, we may be able to rationally design spatially uniform conditions that lead to robust outcomes. In either approach, what are the engineering limits for tissue size and number of different cell types? Our quantitative understanding of spontaneous symmetry breaking is crucial for our ability to predict and control organoid self-organization.

## Acknowledgements

The authors thank the members of Elly Tanaka, Frank Jülicher, and Jan Brugues' research groups and Arghyadip Mukherjee for helpful discussions. KI is supported by the ELBE fellowship from the Center of Systems Biology Dresden. EMT's research is supported by BMBF Systems Microscopy and the DFG FZT111/EXC168.

## Conflict of Interest Statement

The authors declare no competing or financial interests.

## Figure Legends

Figure 1. Organoid models that exhibit spontaneous symmetry breaking and pattern formation. (Left) Cortical organoids show rostral-caudal polarization, segregating the tissue into rostral Six3+ (yellow) and caudal Fgf5+/Irx3+ (blue) regions [18]. In larger tissues, a Turing type mechanism predicts repeating patterns, while a "wave-pinning" model predicts polarization failure due to a uniform fate. (Center) 3D gastruloids give rise to a localized population of T/Brachury+ cells (red), which collectively protrude from the tissue reminiscent of the posterior tail bud [29,30]. Multiple T/Brachury+ protrusions are observed

in larger aggregates. (Right) Neural tube organoids harbor a fluid-filled lumen [41]. Retinoic acid induces the expression of the floor plate marker FoxA2 (orange), whose initial expression shows high cell-to-cell variability (our preliminary observations), but gradually becomes spatially refined to break dorsal-ventral symmetry.

| cortical organoid | 3D gastruloid | neural tube organoid |
|---|---|---|
| 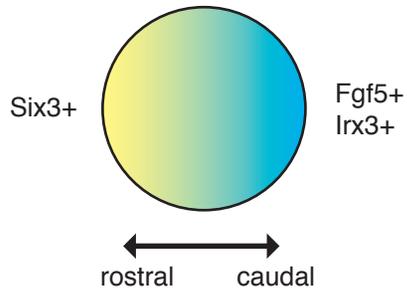 | 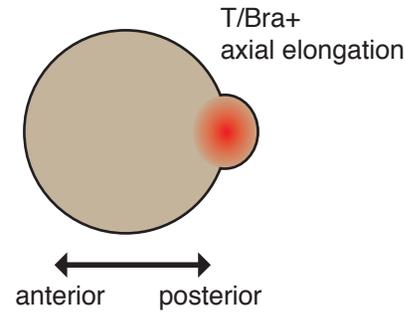 | 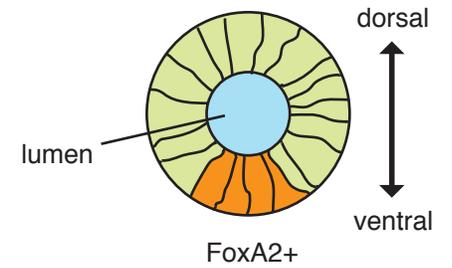 |

tissue size: 300-400 µm | 100-300 µm | 50-100 µm

uniform external cue: none | Wnt activation | all-trans retinoic acid (RA)

pattern in large tissues:

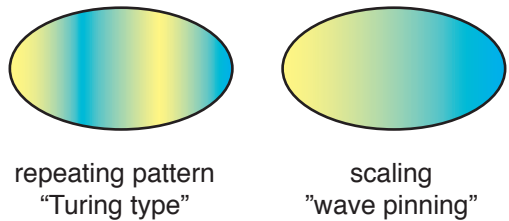 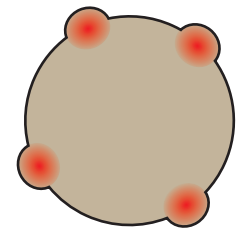 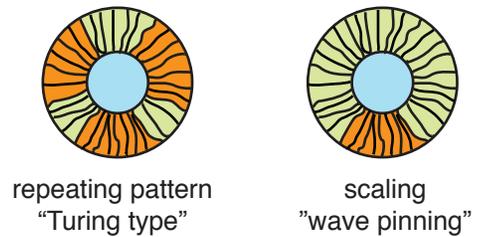

model predictions | multiple axis (reported) | model predictions

future questions:

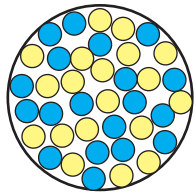 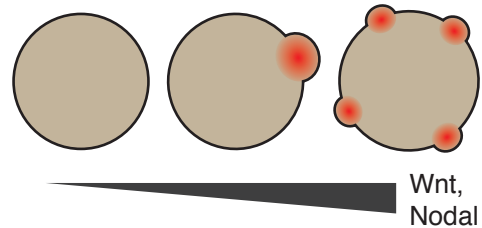 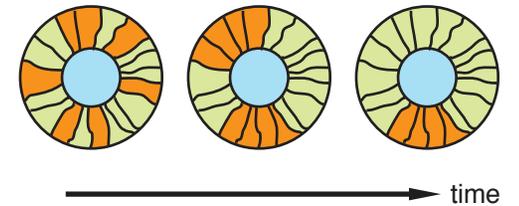

How inhibition of Wnt antagonists results in salt and pepper pattern. | How the level of uniform signals lead to different patterns. | Understanding FoxA2+ dynamics at single cell resolution.